# Investigation of three-body Förster resonance for various spatial configurations of the three interacting Rubidium Rydberg atoms

I.I. Ryabtsev [a, b, *], I.N. Ashkarin [c], I.I. Beterov [a, b, d], E.A. Yakshina [a, b, d], D.B. Tretyakov [a], V.M. Entin [a], P. Cheinet [c]

[a] *Rzhanov Institute of Semiconductor Physics SB RAS, Pr. Lavretyeva 13, 630090, Novosibirsk, Russia*
[b] *Novosibirsk State University, Ul. Pirogova 2, 630090, Novosibirsk, Russia*
[c] *Université Paris-Saclay, CNRS, Laboratoire Aimé Cotton, 91405 Orsay, France*
[d] *Institute of Laser Physics SB RAS, Pr. Lavrentyeva 15b, 630090, Novosibirsk, Russia*

\* *E-mail: ryabtsev@isp.nsc.ru*

Three-body Förster resonances controlled by a dc electric field are of interest for the implementation of three-qubit quantum gates with single atoms captured in optical traps and laser-excited into strongly interacting Rydberg states. In Ref. [P. Cheinet et al., Quantum Electronics **50**(3), 213 (2020)], we proposed and analyzed a new type of three-body Förster resonance $3 \times nP_{3/2} \to nS_{1/2} + (n+1)S_{1/2} + nP_{1/2}$ that can be realized with Rb Rydberg atoms for an arbitrary principal quantum number *n*. Its peculiarity is that the third atom goes into a state with a total angular moment *J*=1/2, which has no Stark structure, so two-body Förster resonances are completely absent. In the present work, an extended theoretical study of this three-body Förster resonance is performed for various spatial configurations of three interacting Rb Rydberg atoms and conditions for their experimental implementation are determined. It was found that one of the resonances has a weak dependence of the resonant electric field on the distance between atoms and is therefore most suitable for performing experiments to observe coherent oscillations of populations of collective three-body states and implement three-qubit quantum gates based on them.

## 1. Introduction

Atoms in highly excited Rydberg states with the principal quantum number *n*>>1 have strong long-range interactions due to large dipole moments, which grow as $n^2$ with increasing *n* [1]. Due to this, the energy of the resonant dipole-dipole interaction of Rydberg atoms grows as $n^4$, and the non-resonant van der Waals interaction as $n^{11}$, and is many orders of magnitude higher than the interaction energy of atoms in low-excited states. This is especially attractive for the creation of quantum computers and simulators with qubits based on single atoms captured in arrays of optical dipole traps [2-5]. The implementation of entangling quantum gates or quantum simulations is achieved by short-term laser excitation of atoms into Rydberg states.

To perform two-qubit quantum gates that create entangled states of qubits, the dipole blockade effect is usually used during laser excitation of Rydberg atoms, when in an ensemble, due to strong non-resonant van der Waals interactions leading to shifts in collective energy levels, it becomes impossible to excite more than one Rydberg atom [6,7]. However, the dipole blockade requires high interaction energies and is therefore realized only for atoms located at small distances from each other (a few microns). This limits the accuracy of spatial laser addressing to individual atoms and the fidelity of quantum gates [3,8].

At the same time, the long-range dipole-dipole interaction of Rydberg atoms remains poorly studied and has not yet been applied to quantum gates. The fundamental feature of such interaction is that it can exhibit resonant and coherent properties. For ordered atoms in arrays of optical dipole traps, the coherent dipole-dipole interaction of Rydberg atoms manifests itself in the form of population oscillations of the collective states of a quasimolecule formed by Rydberg atoms with a fixed distance between them. Such oscillations for two Rb Rydberg atoms were first observed in [9-11] in spatially separated optical dipole traps for two-body Förster resonances tuned by an electric field due to the Stark effect. The study of coherent dipole-dipole interaction (with periodic oscillations of populations and phases of collective states) is valuable for obtaining entangled states of atoms located at relatively large distances from each other (tens of microns). This creates new opportunities for both the implementation of quantum gates and the control of many-body collective states of atoms in quantum simulations.

Population oscillations in the van der Waals interaction were observed in [12] for Rb Rydberg atoms and in [13] for K Rydberg atoms, where individual atoms were captured in arrays of optical dipole traps with a controlled spatial configuration. In [14], the regime of fast population oscillations for two Rb Rydberg atoms in optical traps was achieved under conditions of a two-body Förster resonance, which corresponded to a resonant dipole-dipole interaction. In [15], population oscillations were also achieved for a large disordered ensemble of Rb Rydberg atoms by using the technique of fast Stark switching of the two-body Förster resonance detuning, similar to the photon echo technique. It should be noted that for some Rydberg states of Rb and Cs atoms, the presence of two-body Förster resonances with a small energy defect is possible even without an electric field, which makes it possible to achieve an enhancement of the dipole blockade effect [16-19].

Since quantum computers are currently far from being perfect for all types of qubits (atoms, ions, photons, superconductors, quantum dots), additional error correction protocols must be used for the full functioning of quantum algorithms. The basic elements of such protocols are three-qubit quantum gates, which also allow for further acceleration of quantum computations themselves. Therefore, for a quantum register on neutral atoms, an urgent task is the experimental implementation of three-qubit quantum gates with Rydberg atoms. To do this, it is necessary to obtain controlled coherent three-body interactions of Rydberg atoms in neighboring optical traps.

Three-qubit quantum gates require the ability to control the interactions of three atoms simultaneously. Since such three-body operators are absent in theory for neutral atoms, it is necessary to use an atomic system that is described by a combination of two-body operators, but in fact reduces to a single three-body operator. Such an operator was first proposed and implemented as an electrically controlled three-body Förster resonance in a large ensemble of cold Cs Rydberg atoms in [20]. We performed a similar experiment for $N$=2-5 Rb Rydberg atoms, where the three-body Förster resonance $3 \times nP_{3/2} \rightarrow nS_{1/2}+(n+1)S_{1/2}+nP_{3/2}*$ was explicitly observed for the first time for three or more atoms, while it was absent for two atoms [21]. In subsequent theoretical works, we investigated the coherence of these three-body Förster resonances [22] and proposed a scheme for implementing the three-qubit Toffoli quantum gates based on them [23].

In the case of Rb Rydberg atoms, one of the disadvantages of the above-mentioned three-body Förster resonances, where the third atom changes the moment projection $M$, is the proximity of the two-body Förster resonance $2 \times nP_{3/2} \rightarrow nS_{1/2}+(n+1)S_{1/2}$, which partially overlaps with the three-body resonance in the scale of the control dc electric field. Another disadvantage is that, due to the specific values of quantum defects and polarizabilities of the Rydberg states $nP$ and $nS$ in Rb atoms, the crossing of collective energy levels in the control electric field, corresponding to the three-body Förster resonances, is possible only for states with the values of the principal quantum number $n \leq 38$. At the same time, to increase the fidelity of quantum computations and simulations, it is necessary to use higher Rydberg states with longer lifetimes and transition dipole moments.

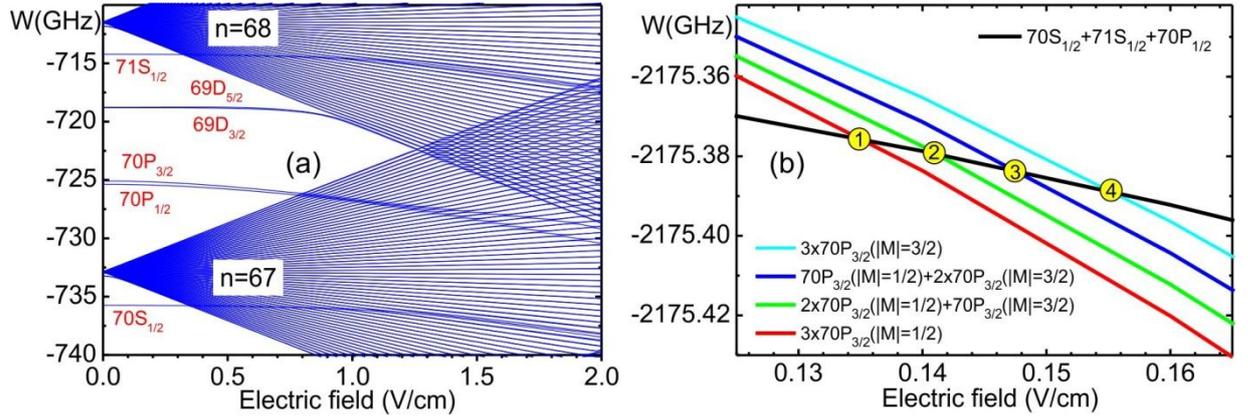

**Fig. 1.** (a) Calculated Stark map of Rydberg states of Rb atoms near the 70*P* state. (b) Calculated Stark structure of the new type of Förster resonance $3\times 70P_{3/2} \to 70S_{1/2}+71S_{1/2}+70P_{1/2}$ for three Rb Rydberg atoms. The intersections of collective states (labeled by numbers) correspond only to three-body Förster resonances, when all three atoms change their states, and two-body resonances are absent [24].

Therefore, in [24] we found a new, simpler three-body resonance $3\times nP_{3/2} \to nS_{1/2}+(n+1)S_{1/2}+nP_{1/2}$, in which there is only one final collective state. As it turned out, such a resonance can be realized for arbitrary initial Rydberg $nP_{3/2}$ states and has no limitation on *n*. Figure 1a shows the calculated Stark map of the Rydberg states of Rb atoms near the 70*P* state, and Figure 1b shows the Stark structure of the Förster resonance of a new type $3\times 70P_{3/2} \to 70S_{1/2}+71S_{1/2}+70P_{1/2}$ for three Rb Rydberg atoms. In this resonance, the intersections of collective states (labeled by numbers) correspond only to three-body Förster resonances, when all three atoms change their states, and two-body resonances are absent at all. Its distinctive feature is that the third atom does not go into a state with a different moment projection *M*, but into a state with a different total moment *J*=1/2, which does not have a Stark structure. Accordingly, the experimental study of such a three-body resonance should be much simpler, since the two-body resonance interfering with its observation is completely absent.

In the subsequent theoretical work, we proposed a scheme for implementing the three-qubit Toffoli quantum gate based on three-body Förster resonances of a new type [25]. Also, a scheme for implementing doubly controlled phase gates *CCΦ* based on these resonances with the addition of a radio-frequency field creating additional Rydberg Floquet levels was developed [26]. Two-body Förster resonances for Rydberg Floquet levels were previously investigated by us experimentally and theoretically in [27,28].

In this paper, an extended theoretical study of a new type of three-body Förster resonance is performed for various spatial configurations of three interacting Rb Rydberg atoms and conditions for their experimental implementation are determined.

## 2. Theory of three-body Förster resonances in Rydberg atoms

A detailed theoretical analysis of three-body Förster resonances $3\times nP_{3/2} \to nS_{1/2}+(n+1)S_{1/2}+nP^{*}_{3/2}$ was performed by us in [22], and three-body Förster resonances of a new type $3\times nP_{3/2}\to nS_{1/2}+(n+1)S_{1/2}+nP_{1/2}$ – in [24].

In Rb Rydberg atoms, ordinary two-body Förster resonances, driven by a weak dc electric field, can be observed for two or more interacting Rb Rydberg atoms. In such resonances, the dipole-dipole interaction causes transitions between the initial state $nP_{3/2}$ and the final state $nS_{1/2}$ or $(n+1)S_{1/2}$ in two atoms, while the other atoms remain in the initial state $nP_{3/2}$, which does not change. Three-body resonances differ from two-body resonances in that the third atom does not remain in the initial state, but changes either the projection of the angular momen *M* or the angular moment *J* itself. Therefore, three-body resonances correspond to transitions in which three interacting atoms change their states simultaneously.

The three-body Förster resonance actually consists of two non-resonant intermediate two-body transitions that occur simultaneously (Fig. 2). The first transition is a normal two-body resonance, while the second transition is due to a non-resonant exchange interaction corresponding to excitation hopping between Rydberg atoms in the *S* and *P* states, with the third atom compensating for the non-zero energy defect of the intermediate Förster resonance. Therefore, three-body resonances are less efficient than two-body resonances with weak dipole-dipole interactions. However, when the three-body resonance is precisely tuned by an electric field, its contribution to the population transfer generally exceeds that of the two-body interaction, which is non-resonant in this case.

Let the initial collective state be state 1 (Fig. 2). The intermediate collective state 2 corresponds to two atoms in *S* states and one atom in the initial *P* state. The final state is state 3, which has the other moment of the atom in the *P* state. The detunings $\Delta_{12}$ and $\Delta_{23}$ are controlled by the electric field. Three-body resonance occurs at $\Delta_{12} = \Delta_{23}$ [22,24].

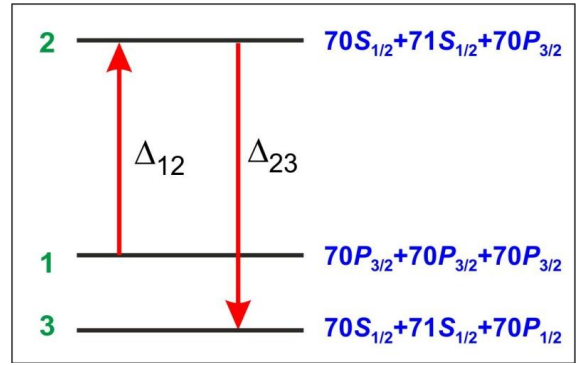

**Fig. 2.** Simplified scheme of the new type three-body Förster resonance $3\times 70P_{3/2} \rightarrow 70S_{1/2}+71S_{1/2}+70P_{1/2}$ for three Rb Rydberg atoms. The initial state is state 1. The intermediate state 2 corresponds to two atoms in *S* states and one atom in the initial *P* state. The final state is state 3, which has the other moment of the atom in the *P* state. The detunings $\Delta_{12}$ and $\Delta_{23}$ are controlled by the electric field. The three-body resonance occurs at $\Delta_{12} = \Delta_{23}$ due to the Stark tuning of the levels by a dc electric field.

In [22], for three frozen Rydberg atoms in a triangular configuration (with the same interaction energies for each pair of atoms), we obtained the following analytical solution for the line shape and time evolution of the amplitude of three-body Förster resonances (the probability of transition from the initial state $nP_{3/2}$ to the final state $nS_{1/2}$ in each atom):

$$\rho_3 \approx \frac{\Omega_0^2/3}{(\Delta-\Delta_0)^2+\Omega_0^2}\sin^2\left[\frac{t}{2}\sqrt{(\Delta-\Delta_0)^2+\Omega_0^2}\right], \quad (1)$$

where $\Delta = \Delta_{12} - \Delta_{23}$ is the detuning from the unperturbed three-body Förster resonance, $\Delta_0 = -2\Omega_{23} + (4\Omega_{23}^2 - 6\Omega_{12}^2)/(\Delta_{23}+2\Omega_{12})$ is the dynamic shift of the three-body resonance due to non-resonant intermediate interactions with the matrix elements of the dipole-dipole interaction operators $\Omega_{12} = V_{12}/\hbar$ and $\Omega_{23} = V_{23}/\hbar$ on the 1→2 and 2→3 transitions, and $\Omega_0 = 4\sqrt{6}\Omega_{12}\Omega_{23}/(\Delta_{23}+2\Omega_{12})$ is the frequency of population oscillations of collective states in the exact resonance. This formula is identical to the Rabi oscillations at a two-photon transition in a three-level system with a detuned intermediate level 2, which is not populated, and the population oscillations occur only between levels 1 and 3. A certain interaction time *t* for long-lived Rydberg states can be specified by the method of Stark switching of levels in a pulsed electric field, as in our works [28,29].

It follows from Eq. (1) that the three-body resonance experiences a dynamic shift consisting of two parts: part with $-2\Omega_{23}$ arises due to always resonant exchange interactions of atoms in the *S* and *P* states, and the other part is a dynamic Stark shift caused by intermediate non-resonant interactions. Therefore, the position of the three-body resonance in the scale of the control electric field depends on the interaction energy and the ratio of the energies of the intermediate transitions $1 \rightarrow 2$ and $2 \rightarrow 3$. In real Rydberg atoms, there is a Zeeman structure of the Rydberg levels, which leads to many interaction channels with different matrix elements of the dipole-dipole interaction. Therefore, due to the dynamic shift, we should observe not single three-body resonances, but a set of individual resonances arising at slightly different values of the electric field. If the field difference is large enough, it is possible to work with individual interaction channels. By choosing the spatial configuration of the three atoms, some channels can be suppressed.

Earlier in [22] we showed that in order to reduce the number of interaction channels of three-body Förster resonances, the optimal configuration of three atoms is their uniform arrangement along the quantization axis *Z*, which is chosen along the direction of the control electric field. In this case, only two well-separated three-body Förster resonances remain, corresponding to two interaction channels due to the fact that in this configuration, predominantly atoms interact that are in states with the same moment projection *M*. This conclusion was confirmed in [24] for three-body Förster resonances of a new type $3 \times nP_{3/2} \rightarrow nS_{1/2} + (n+1)S_{1/2} + nP_{1/2}$.

However, such a configuration may not always be convenient for conducting experiments. Therefore, in this paper, we have carried out an extended theoretical study of the three-body Förster resonance of a new type for various spatial configurations of three interacting Rb Rydberg atoms and determined the conditions for their experimental implementation.

### 3. Results of numerical simulations for three spatial configurations

Accurate analytical calculations for an arbitrary geometry of the arrangement of three interacting Rydberg atoms are impossible, so we performed numerical simulations of the probability amplitudes of all collective states based on the Schrödinger equation in the same way as was done in [22] for three-body Förster resonances $3 \times 37P_{3/2} \rightarrow 37S_{1/2} + 38S_{1/2} + 37P^*_{3/2}$. A complete model of the interaction of atoms was used, taking into account the Zeeman structure of Rydberg levels. To simplify the calculations, collective states of the atomic system with an energy defect of more than 2 GHz in zero electric field were excluded from consideration. Thus, for the three-body Förster resonance in atoms in the initial state $70P_{3/2}(M=1/2)$, a complete calculation required taking into account 360 collective states with all possible values of the moment projections. Each collective state was a product of the states of three atoms in the basis states $70S_{1/2}(M = \pm 1/2)$, $71S_{1/2}(M = \pm 1/2)$, $70P_{1/2}(M = \pm 1/2)$, $70P_{3/2}(M = \pm 1/2, \pm 3/2)$.

The final radiative lifetimes of all Rydberg states, calculated by us according to the work [30] taking into account the influence of background thermal radiation at T=300 K (70*S* – 152 μs; 71*S* – 156 μs; 70*P*$_{1/2}$ – 189 μs; 70*P*$_{3/2}$ – 191 μs), were also phenomenologically taken into account by introducing a weak quenching of the probability amplitudes into the Schrödinger equation. Although this leads to non-conservation of the total initial population of the collective states, this procedure allows us to calculate the maximum possible contrast of population oscillations for the implementation of three-qubit quantum gates [23,25,26].

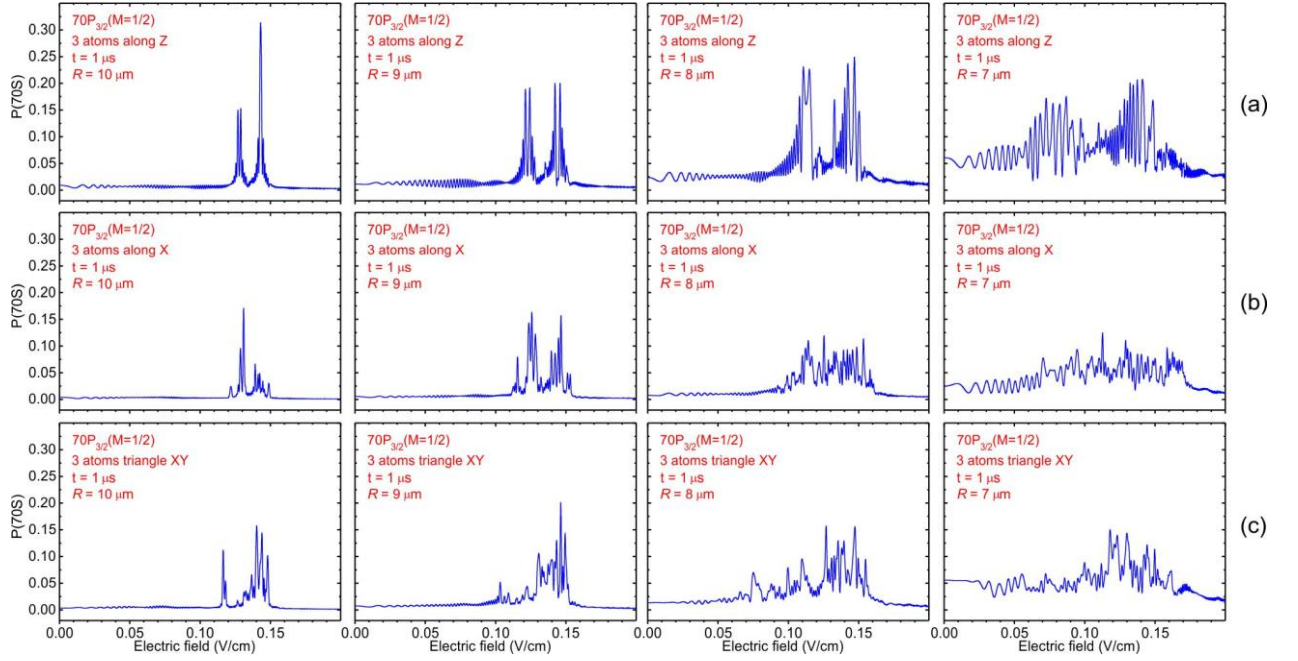

**Fig. 3.** Results of numerical simulations of the three-body Förster resonance of a new type $3\times 70P_{3/2}(M=1/2)\rightarrow 70S_{1/2}+71S_{1/2}+70P_{1/2}$ for three frozen Rydberg atoms Rb in several spatial configurations at an interaction time of 1 μs. (a) Three atoms are uniformly distributed along the $Z$ axis with an interatomic distance of $R = 7$-$10$ μm. (b) The same for the arrangement along the $X$ axis. (c) Three atoms are located in the $XY$ plane in the form of an equilateral triangle with an edge of $R = 7$-$10$ μm.

The calculations performed showed that for three Rydberg atoms in the initial state $70P_{3/2}(M=1/2)$, uniformly distributed along the Z axis with the distance between the atoms $R=10$ μm at an interaction time of 1 μs, there is a relatively weak three-body interaction, when the three-body resonances are not broadened (Fig. 3a). As expected, in this configuration only two resonances arise, which correspond to only two interaction channels due to the fact that in this configuration, predominantly atoms interact that are in states with the same moment projection $M$. Their resonant electric fields of 0.128 and 0.143 V/cm are close to the calculated value of 0.135 V/cm for crossing collective levels (Fig. 1b) in the absence of interaction, taking into account additional dynamic shifts. The resonance amplitudes tend to the maximum possible value of 1/3, and their width when converted to the frequency scale corresponds to the Fourier width of the interaction pulse (about 1 MHz). The resonances are well resolved, and by adjusting the electric field, a specific channel of three-body interaction can be selected.

As the distance between atoms decreases (Fig. 3a), the effective energy of three-body interaction, which depends on the distance as $R^{-6}$, increases significantly. As a result, the calculated resonances begin to broaden noticeably, shift, and partially overlap in the presence of population oscillations. From the calculated dependences of the shifts of the centers of two channels of the three-body Förster resonance on the distance between atoms $R$ (Fig. 4a), it can be seen that one of the resonances has a weaker dependence on $R$. It can be associated with the interaction of atoms in states with the same moment projection $M$. The weak dependence on $R$ is due to the fact that the dipole moments of the up and down transitions from the initial state are almost identical (the radial parts are 4954 and 5082 a.u., respectively), so the dynamic shifts of the collective states compensate each other. The other resonance quickly shifts toward a lower electric field, and its wing has a nonzero value even in zero field. It can be associated with the interaction of atoms in states with different moment projections $M$, for which the dipole moments of the up and down transitions, and, consequently, the dynamic shifts of the collective states, are very different. The observed oscillations of the populations on the wings of the three-body resonance have a period that increases with decreasing electric field, which is explained by the quadratic nature of the Stark effect.

The three-body Förster resonance with a weak dependence on $R$ is more preferable for performing three-qubit quantum gates, since fluctuations in the interatomic distance will have a lesser effect on the fidelity of such gates. Calculations have shown that when tuning to such a three-body resonance in an electric field of 0.1431 V/cm for an interatomic distance of $R$=10 μm, oscillations of the population of the initial collective state $3\times70P_{3/2}(M=1/2)$ occur with a contrast exceeding 95% (Fig. 4b) and a frequency of $\Omega$=1.51 MHz. It is noteworthy that all the oscillation minima are close to 0, that is, the population completely goes to other collective states. At the same time, the amplitude of the maxima gradually decreases, that is, part of the population does not return to the initial state. The limiting contrast of oscillations and their damping are determined mainly by the finite radiative lifetime of Rydberg states. Nevertheless, the presence of sufficiently contrasting oscillations allows us to consider them as a basis for three-qubit quantum gates, where it is required to have a controlled phase incursion of the collective state over a certain interaction time.

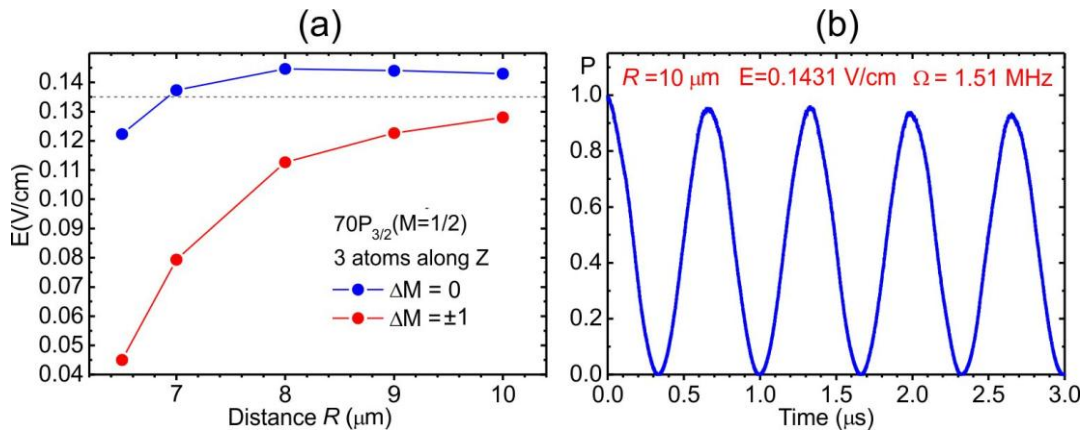

**Fig. 4.** (a) Dependences of the shifts of the centers of two channels of the three-body Förster resonance of a new type $3\times70P_{3/2}(M=1/2)\rightarrow70S_{1/2}+71S_{1/2}+70P_{1/2}$ in Fig. 3a on the distance $R$ between atoms. The dotted line shows the calculated position of the resonance in Fig. 1b without taking into account the dynamic shifts. (b) Population oscillations of the initial collective state $3\times70P_{3/2}(M=1/2)$ at the distance between atoms $R$ = 10 μm and tuning the electric field to the three-body resonance in an electric field of 0.1431 V/cm. The contrast of the oscillations exceeds 95%, which allows them to be considered as a basis for three-qubit quantum gates.

We also performed numerical simulations of the three-body Förster resonance $3\times70P_{3/2}(M=1/2)\rightarrow70S_{1/2}+71S_{1/2}+70P_{1/2}$ in two other spatial configurations: a one-dimensional chain along the $X$-axis (Fig. 3b) and an equilateral triangle in the $XY$ plane (Fig. 3c) for the same different distances between atoms. As expected, many additional interaction channels of Rydberg atoms are included in these configurations without specific selection rules. Even at the maximum distance $R$=10 μm, there are many overlapping resonances in the region of 0.14 V/cm. At the same time, for the orientation of atoms along the $X$-axis, a separate resonance is observed at 0.131 V/cm, and in the triangular configuration such a resonance occurs at 0.116 V/cm. However, the position of these resonances turns out to be strongly dependent on $R$, and already when $R$ decreases to 9 μm, they noticeably shift and broaden. It can be concluded that in these configurations, observation of a single channel of the three-body Förster resonance is possible only for $R$>9 μm, and the resonance amplitude turns out to be noticeably smaller than for atoms in the configuration along $Z$. In general, numerical simulations confirmed that in these configurations, high-contrast population oscillations are impossible, so they are of no interest for further applications in quantum information processing, although they are experimentally observable.

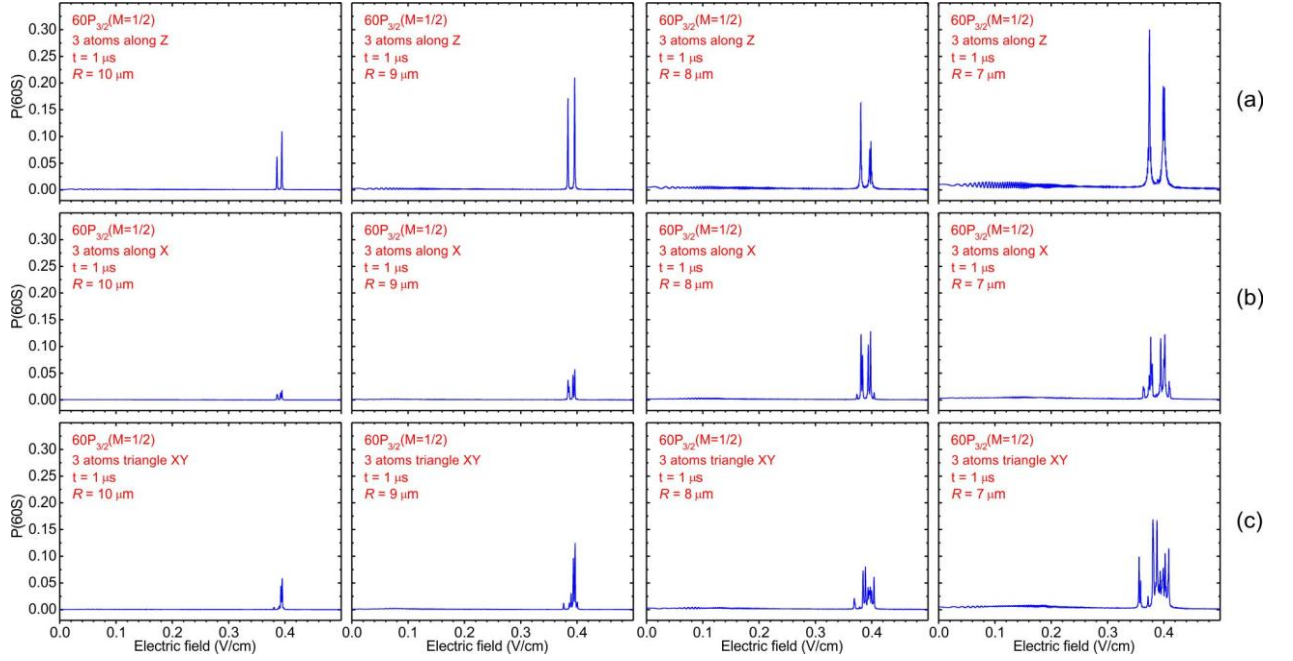

**Fig. 5.** Results of numerical simulations of the three-body Förster resonance of a new type $3\times 60P_{3/2}(M=1/2)\to 60S_{1/2}+61S_{1/2}+60P_{1/2}$ for three frozen Rb Rydberg atoms in several spatial configurations with an interaction time of 1 μs. (a) Three atoms are uniformly distributed along the Z axis with an interatomic distance of $R = 7\text{-}10$ μm. (b) The same for the arrangement along the X axis. (c) Three atoms are located in the XY plane in the form of an equilateral triangle with an edge of $R = 7\text{-}10$ μm.

The three-body Förster resonance $3\times 70P_{3/2}(M=1/2)\to 70S_{1/2}+71S_{1/2}+70P_{1/2}$ occurs in an electric field of about 0.135 V/cm. For experimental observation of such a resonance, it is necessary to have a scanning step of the electric field of ~0.1 mV/cm, and the parasitic electric fields, which are always present in experiments, must be much smaller than the magnitude of the resonant field.

In case of detection of parasitic fields, experiments can be performed with lower Rydberg states, for which the resonant electric field has a greater value. Therefore, we also performed numerical calculations of the three-body Förster resonance of a new type $3\times 60P_{3/2}(M=1/2)\to 60S_{1/2}+61S_{1/2}+60P_{1/2}$ for three Rb Rydberg atoms in the same spatial configurations (Fig. 5). Although the energy of the three-body interaction for this resonance will be lower due to smaller values of the transition dipole moments and large detunings of the intermediate levels, its resonant electric field is near 0.38 V/cm, which can allow experiments to be performed even in the presence of parasitic fields up to 0.1-0.2 V/cm. From Fig. 5 it can be seen that the rather weak three-body interaction of atoms in the $60P_{3/2}$ state compared to atoms in the $70P_{3/2}$ state gives significantly narrower three-body resonances for any spatial configuration, so the experiments can also be performed for any configuration. At the same time, the observation of high-contrast population oscillations should be expected only for the configuration along Z, the distance between atoms of 7 μm and in an electric field of 0.381 or 0.398 V/cm.

The numerical simulations performed above required considerable computer time (4-5 hours per graph) due to the necessity of taking into account a large number of collective states and projections of magnetic moments of Rydberg levels. In such a situation, it is practically impossible to perform calculations that require large averaging to take into account fluctuations of interatomic distances, which always occur in real experiments with atoms in optical traps. To perform such calculations, we have built a simplified theoretical model in which the signs of the moment projections were not taken into account (i.e., the simplified model was built not for Zeeman, but for Stark Rydberg sublevels). Its functionality was previously tested in calculations of two-body Förster resonances for Rb Rydberg atoms in the states $36P_{3/2}$, $37P_{3/2}$, $39P_{3/2}$, which demonstrated satisfactory agreement between theory and experiment in the positions and amplitudes of Förster resonances in disordered ensembles [27,28].

When comparing the calculation results in the simplified and full models for three atoms uniformly spaced along the Z axis with an interaction time of 1 μs, their satisfactory agreement was confirmed for interatomic distances of 10 μm (Fig. 6a, b) and 8 μm (Fig. 6c, d). In both cases, only two resonances were observed in this configuration. At $R = 10$ μm, the simplified model yields resonance fields of 0.125 and 0.140 V/cm (Fig. 6a), while the full model yields 0.128 and 0.143 V/cm (Fig. 6b). Their slight difference is due to some difference in the interaction energy given by the two models. Also, as a result, the population oscillations at exact resonances demonstrate different final phases in the simplified and full models. Nevertheless, an important conclusion is that both models predict the presence of population and phase oscillations of collective three-body states, which can be used to perform three-qubit quantum gates. Accurate calculations of such gates should be performed in the full model.

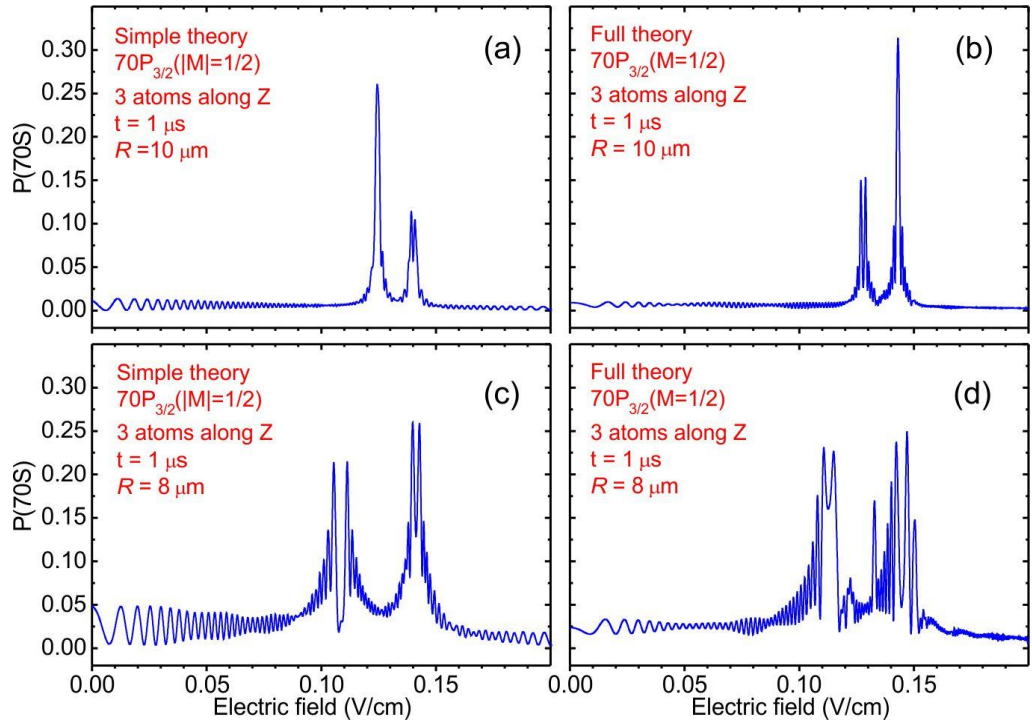

**Fig. 6.** Results of numerical simulations of the spectra of the three-body Förster resonance of a new type $3 \times 70P_{3/2}(M=1/2) \rightarrow 70S_{1/2}+71S_{1/2}+70P_{1/2}$ for three Rb Rydberg atoms in a linear spatial configuration along the Z axis with a distance R between atoms at an interaction time of 1 μs. (a) Results of a simplified theoretical model at $R=10$ μm. (b) Results of a full theoretical model at $R=10$ μm. (c) Results of a simplified theoretical model at $R=8$ μm. (d) Results of a full theoretical model at $R=8$ μm.

Based on a simplified model, we calculated the case when three atoms in a linear spatial configuration along the Z axis with an average distance between atoms $R = 10$ μm had random fluctuations of this distance. Figures 7a-g show the results of numerical calculations of the spectra of the three-body Förster resonance of a new type $3 \times 70P_{3/2}(|M|=1/2) \rightarrow 70S_{1/2}+71S_{1/2}+70P_{1/2}$ with an interaction time of 0.4 μs and averaging over 1000 random fluctuations of the distance in the ranges: (a) $\Delta R=\pm 0$ μm; (b) $\Delta R=\pm 1$ μm; (c) $\Delta R=\pm 2$ μm; (d) $\Delta R=\pm 4$ μm. In the absence of fluctuations (Fig. 7a), the two peaks of the Förster resonance had a minimum width determined by the Fourier width of the interaction pulse, and their amplitudes were close to the maximum possible value of 1/3. From Fig. 4a, we already know that the right peak at an electric field strength of 0.14 V/cm is more stable to a change in $R$ than the left peak at 0.124 V/cm. This was confirmed by introducing fluctuations $\Delta R$ in Fig. 7b-d. While the left peak begins to noticeably broaden and decrease in amplitude as $\Delta R$ increases, the right peak remains narrow up to fluctuations $\Delta R = \pm 4$ μm, and at $\Delta R = \pm 1$ μm its amplitude is still close to the maximum. This means that this peak retains coherence even in the presence of noticeable fluctuations in $R$.

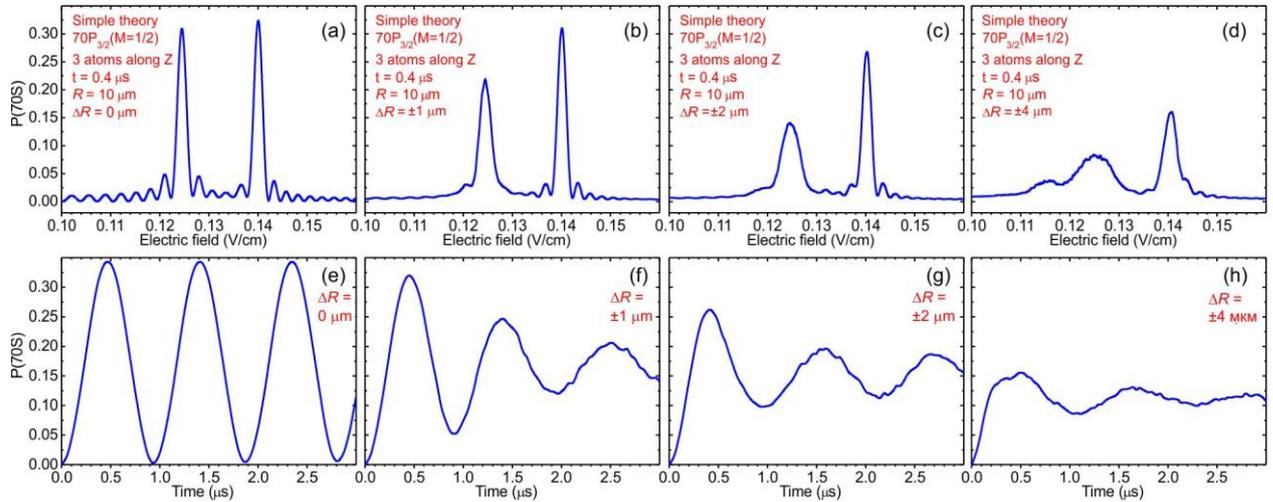

**Fig. 7.** (a)-(d) Results of numerical simulations in a simplified theoretical model of the spectra of the three-body Förster resonance of a new type $3\times 70P_{3/2}(|M|=1/2) \rightarrow 70S_{1/2}+71S_{1/2}+70P_{1/2}$ for three Rb Rydberg atoms in a linear spatial configuration along the $Z$ axis with an average distance between atoms $R=10$ μm for an interaction time of 0.4 μs and averaging over 1000 random fluctuations of the distance in the range: (a) $\Delta R=\pm 0$ μm; (b) $\Delta R=\pm 1$ μm; (c) $\Delta R=\pm 2$ μm; (d) $\Delta R=\pm 4$ μm. (e)-(h) The same for population oscillations when tuning to a three-body resonance in an electric field of 0.14 V/cm.

The preservation of its coherence was confirmed by numerical calculation of population oscillations when tuning to a three-body resonance in an electric field of 0.14 V/cm. At $\Delta R=\pm 0$ μm, the contrast of these oscillations is close to 100% (Fig. 7e). At $\Delta R=\pm 1$ μm, the amplitude of the first oscillation decreases to 96% and oscillation damping appears with a coherence time of about 2 μs (Fig. 7f). Nevertheless, such oscillations should be easily observable in the experiment. Moreover, even at $\Delta R=\pm 2$ μm (Fig. 7g) and $\Delta R=\pm 4$ μm (Fig. 7h), population oscillations can still be observed. In real experiments with single atoms in optical dipole traps, the temperature of the atoms does not exceed 10 μK, which corresponds to fluctuations in the distances between atoms of less than 0.1 μm. With such small fluctuations, our calculations showed that the coherence time increases to 25 μs, so coherent population oscillations, and therefore precise three-qubit quantum gates, should be quite feasible for the three-body Förster resonance of the new type.

## 4. Conclusion

In our previous experiments [21], three-body Förster resonances were studied under conditions of a disordered atomic ensemble, which led to a significant broadening of the resonances and the disappearance of coherence due to the uncertain energy of the dipole-dipole interaction.

For ordered atoms in optical dipole trap arrays, the coherent dipole-dipole interaction of Rydberg atoms manifests itself as population oscillations of collective states [9–11, 14, 15]. The frequency and contrast of the oscillations depend on the energy of the dipole-dipole interaction (the distance between the atoms and their spatial configuration) and the energy defect of the Förster resonance, determined by the structure of the energy levels and the external control electric field. For three Rydberg atoms, such coherent population oscillations have not yet been observed experimentally under the conditions of the three-body Förster resonance, although they have been observed for three Rydberg atoms of Rb [12] and K [13] in the van der Waals interaction, as well as in three-body interactions in a large ensemble of Rb Rydberg atoms near the energy-detuned two-body Förster resonance [31].

Theoretical calculations of three-body Förster resonances of a new type $3 \times nP_{3/2} \rightarrow nS_{1/2} + (n+1)S_{1/2} + nP_{1/2}$ for various spatial configurations and various Rydberg states confirmed the possibility of their experimental observation. For this, three Rydberg atoms should be uniformly arranged along the quantization axis $Z$, specified by the control electric field, which adjusts the collective energy levels to exact three-body resonances. Despite the complex structure of the atomic levels, in this configuration only two interaction channels of atoms are observed, which manifest themselves in the form of two three-body resonances close in the electric field.

It was found that one of these resonances has a weak dependence of the resonant electric field on the distance between atoms and retains coherent properties even at relatively large fluctuations of the interatomic distance, which are always present in experiments with single atoms in optical dipole traps. The weak dependence is due to the fact that at this resonance the selection rule for conservation of the moment projection of Rydberg levels is fulfilled, which leads to suppression of the dynamic shift of the three-body resonance. Therefore, this resonance is the most suitable for performing experiments on observing coherent population oscillations of collective three-body states and implementing three-qubit quantum gates based on them.

We also note that many-body electrically controlled Förster resonances for large ensembles of Rydberg atoms were studied experimentally and theoretically in [32–34], where the possibility of observing four-body and higher resonances was noted, requiring, however, significantly higher interaction energies.

This work was supported by the Russian Science Foundation (Project 23-12-00067, https://rscf.ru/project/23-12-00067/).